\documentclass[conference, final]{IEEEtran}
\IEEEoverridecommandlockouts
\usepackage{cite}
\usepackage{amsmath,amssymb,amsfonts}
\usepackage{algorithmic}
\usepackage{graphicx}
\usepackage{textcomp}

\usepackage[inline]{enumitem}
\usepackage{algorithmic}
\usepackage{graphicx}
\usepackage{textcomp}
\usepackage{xcolor}

\usepackage{witharrows}

\usepackage[flushleft]{threeparttable}
\usepackage{tablefootnote}
\usepackage{array}

\newcolumntype{P}[1]{>{\centering\arraybackslash}m{#1}}
\usepackage{multicol}
\usepackage{multirow}
\usepackage{tabulary}
\usepackage{colortbl}
\definecolor{mygray}{gray}{0.90}
\usepackage{makecell}
\usepackage{booktabs}
\usepackage{subcaption}
\usepackage{stmaryrd}
\usepackage{float}
\usepackage{pifont}

\usepackage{arydshln}
\colorlet{mygray}{gray!15!white}

\usepackage[switch,columnwise]{lineno}

\usepackage{url,hyperref,microtype}
\hypersetup{
    colorlinks=true,
    citecolor=blue,
    linkcolor=blue,
    filecolor=magenta,      
    urlcolor=black,
}

\def\BibTeX{{\rm B\kern-.05em{\sc i\kern-.025em b}\kern-.08em
    T\kern-.1667em\lower.7ex\hbox{E}\kern-.125emX}}
\begin{document}

\title{A Mental Health Monitoring Framework via EEG-Based Mixed Emotion Recognition\\
}

\author{

\IEEEauthorblockN{Stefanos Gkikas}
\IEEEauthorblockA{\textit{Honda Research Institute Japan} \\
Wako City, Japan \\
stefanos.gkikas@jp.honda-ri.com}

\and

 \IEEEauthorblockN{Christian Arzate Cruz}
 \IEEEauthorblockA{\textit{Honda Research Institute Japan} \\
 Wako City, Japan  \\
 christian.arzate@jp.honda-ri.com}

 \and

\IEEEauthorblockN{Yu Fang}
\IEEEauthorblockA{\textit{Honda Research Institute Japan} \\
Wako City, Japan  \\
yu.fang@jp.honda-ri.com}

\and

\IEEEauthorblockN{Eric Nichols}
\IEEEauthorblockA{\textit{Honda Research Institute Japan} \\
Wako City, Japan \\
e.nichols@jp.honda-ri.com}

\and

\IEEEauthorblockN{Guangliang Li}
\IEEEauthorblockA{\textit{Faculty of Information Science}
\textit{and Engineering} \\
\textit{Ocean University of China}\\
Qingdao, China \\
guangliangli@ouc.edu.cn}

\and

\IEEEauthorblockN{Zongyi Jiang}
\IEEEauthorblockA{
\textit{University of Canberra} \\
Canberra, Australia \\
zongyi.jiang@canberra.edu.au
}

\and

\IEEEauthorblockN{Raul Fernandez Rojas}
\IEEEauthorblockA{\textit{BioSIS (Biosensing \& Intelligent Systems) Lab} \\
\textit{Centre for Intelligent Computing and Systems} \\
\textit{University of Canberra} \\
Canberra, Australia \\
raul.fernandezrojas@canberra.edu.au}

\and

\IEEEauthorblockN{Zacharias Kamarianakis}
\IEEEauthorblockA{\textit{Department of Electronic Engineering} \\\
\textit{Hellenic Mediterranean University}\\
Chania, Greece \\
zkamar@hmu.gr}

\and

\IEEEauthorblockN{Giorgos Giannakakis}
\IEEEauthorblockA{\textit{Department of Electronic Engineering} \\
\textit{Hellenic Mediterranean University}\\
Chania, Greece \\
ggian@hmu.gr}

}

\maketitle

\begin{abstract}
Mental health monitoring relies on accurately characterizing emotional states as key indicators of psychological well-being. These states range from positive and negative affect to mixed conditions, in which both are experienced simultaneously. Automatic recognition systems are essential for enabling continuous and passive monitoring in real-world settings. EEG is a promising modality for affective state recognition, yet its application to mixed emotions remains underexplored, with most studies targeting binary valence or standard discrete categories.
This study proposes an EEG-based framework for mixed emotion recognition and evaluates it on the Dataset for Mixed Emotion Recognition (DMER), which includes data from $73$ participants across positive, negative, and mixed affective categories. The framework is assessed under binary and three-class classification settings using a subject-independent protocol, together with a systematic analysis of individual channel contributions and regional electrode clusters. The best results reach $70.09\%$ for the binary task and $49.89\%$ for the three-class task, both exceeding their corresponding chance levels. 
The central electrode group (C3, C4, Cz) provides performance comparable to the full $18$-channel configuration in both classification settings despite utilizing only $3$ electrodes.
The channel-level analysis further shows a lateralization shift between tasks, with right-posterior emphasis for binary emotion recognition and left-posterior emphasis for three-class recognition, suggesting that mixed emotion recognition relies on partially distinct spatial EEG information. 
These findings establish a deep learning reference point for EEG-based mixed emotion recognition and support the feasibility of reduced-channel EEG configurations for passive mental health monitoring.

\end{abstract}

\begin{IEEEkeywords}
Electroencephalogram, emotion assessment, affective computing, attention
\end{IEEEkeywords}


\section{Introduction}
Mixed emotions refer to the simultaneous co-occurrence of positive and negative affective components within a single emotional episode \cite{larsen_mcgraw_2011}. They are not a midpoint on a valence continuum but a distinct affective category requiring independent assessment of both affective poles \cite{oh_tong_2022}. Subjective instruments such as the Positive and Negative Affect Schedule \cite{watson_clark_1988} and valence-arousal-dominance ratings \cite{bradley_lang_1994} remain common tools for measuring affective experience. The conflicting affective components of mixed states are difficult to reconstruct accurately after the fact, limiting retrospective self-report when affective states must be monitored continuously over time \cite{aaker_drolet_2008}.

Mental health disorders represent a major global health burden, with the COVID-19 pandemic estimated to have added $53$ million cases of major depressive disorder and $76$ million cases of anxiety disorders globally in a single year \cite{santomauro_herrera_2021}. Mixed affective states are a meaningful dimension of emotional experience because they capture simultaneous positive and negative components that standard single-valence descriptions cannot adequately represent \cite{larsen_mcgraw_2014}. Research in clinical populations undergoing psychotherapy has shown that concurrent positive and negative affect prospectively relates to improvements in psychological well-being, supporting the relevance of mixed emotional experience for mental health monitoring \cite{adler_hershfield_2012}.

The limitations of self-report are particularly relevant in settings where continuous and unobtrusive monitoring is required. Electroencephalogram (EEG) frequency-band features have been studied across psychiatric conditions, including depression, bipolar disorder, anxiety, and post-traumatic stress disorder \cite{newson_thiagarajan_2019}. Wearable artificial intelligence applied to physiological signals has demonstrated promise for passive, automated monitoring of anxiety and depression, motivating the development of continuous mental health monitoring systems \cite{abd_alrazaq_alsaad_2023}. Among physiological modalities, EEG captures neural correlates of affective dynamics at millisecond temporal resolution, providing access to central nervous system activity that peripheral biosignals cannot directly measure \cite{alarcao_fonseca_2019}.

Automated emotion recognition from EEG has advanced substantially, with deep learning methods achieving strong performance on discrete emotion classification tasks \cite{song_zheng_2020}. Mixed affective states have received substantially less attention from a computational standpoint. The dataset introduced by Yang et al. \cite{yang_liu_2024} is among the few publicly available resources specifically designed for mixed emotion recognition, providing EEG, galvanic skin response (GSR), photoplethysmography (PPG), and facial video from $73$ participants across positive, negative, and mixed affective categories. 
In this work, we investigate EEG-based mixed-emotion recognition through a unified deep-learning framework, enabling a systematic analysis of individual electrodes and regional EEG groups across binary and three-class affective state classification.

\section{Related Work}
\label{related_work}

Automated emotion recognition has been studied across a broad range of physiological and behavioral modalities, including electrodermal activity, photoplethysmography, cardiovascular signals, facial video, and body movements \cite{koelstra_muhl_2012, cruz_gkikas_emotion_acii_2026}. EEG occupies a distinct position among these modalities since it captures central nervous system correlates of affective states at millisecond resolution, encoding neural dynamics that peripheral biosignals cannot directly reflect \cite{alarcao_fonseca_2019}.

Multimodal approaches that combine physiological and behavioral signals have been explored to capture complementary aspects of the affective response. EEG combined with facial expression analysis has been applied to continuous emotion detection, with results showing that the contribution of each modality varies across tasks and evaluation settings \cite{soleymani_asghari_2016}. Deep feature-level fusion of EEG trials and facial video sequences has shown improvements over unimodal baselines in multi-class emotion classification \cite{muhammad_hussain_2023}. These results highlight that increases in fusion performance are task- and architecture-dependent and come at the cost of greater sensor requirements and deployment complexity.
EEG-based emotion-recognition pipelines have combined hand-crafted spectral features with deep neural architectures, including deep belief networks trained on differential entropy features \cite{zheng_lu_2015}. Differential entropy computed across the delta, theta, alpha, beta, and gamma frequency bands became a widely used feature representation for EEG emotion classification \cite{duan_zhu_2013}. Graph convolutional networks were subsequently applied to model the topology of EEG electrodes, learning inter-channel functional relationships that static feature representations cannot capture \cite{song_zheng_2020}. Transformer-based decoders have further advanced this direction by combining convolutional modules for local temporal and spatial feature extraction with self-attention modules that model global temporal dependencies in EEG sequences \cite{song_zheng_2023}.
Despite these advances, most EEG-based emotion recognition studies have focused on discrete categories or dimensional representations such as valence, arousal, and dominance. Recent work has extended this to emotion distribution learning via multimodal fusion of physiological and behavioral signals \cite{li_liu_2026}, yet mixed emotions, defined as the simultaneous co-occurrence of positive and negative affect that cannot be reduced to a single affective state \cite{larsen_mcgraw_2011}, remain computationally underexplored.


\section{Methodology}
\label{sec:methodology}
This section describes the proposed framework for EEG-based mixed emotion recognition. We first detail how raw EEG signals are represented as input tokens, then describe the asymmetric attention architecture used for classification, and finally outline the training procedure and loss formulation.

\subsection{Signal Representation}

Each EEG trial is represented as a raw multi-channel waveform. All temporal segments within a trial are concatenated along the time axis, resulting in a single input tensor:
\begin{equation}
\mathbf{X} \in \mathbb{R}^{B \times C \times L}, \qquad
L = \left\lfloor L_0 \,/\, s \right\rfloor,
\end{equation}
where $B$ is the batch size, $C$ is the number of selected EEG channels, $L_0$ is the total trial length after concatenation, and $s$ is the temporal stride.
Specifically, the temporal stride $s$ subsamples the raw $300$~Hz EEG waveform by retaining every $s$-th temporal sample; therefore, $s=1$ preserves the original sampling rate, while, for example, $s=10$ corresponds to an effective sampling rate of $30$~Hz.
No hand-crafted feature extraction is applied; the model operates directly on the waveform values. Geometric information is incorporated by encoding each normalized time position $t \in [-1, 1]$ using Fourier features:
\begin{equation}
\begin{split}
\gamma(t) = \bigl[&\sin(\pi s_1 t),\ \cos(\pi s_1 t),\ \ldots,\\
                  &\sin(\pi s_K t),\ \cos(\pi s_K t),\ t\bigr],
\end{split}
\end{equation}
where $\{s_k\}_{k=1}^K$ span $[1, f_{\max}/2]$. 
We employ $K=6$ frequency bands and $f_{\max}=10$ throughout all experiments.
The time axis is flattened into a sequence of $L$ tokens, with waveform channels and positional features concatenated at each step to form the token matrix:
\begin{equation}
\mathbf{T} \in \mathbb{R}^{B \times L \times C'},
\qquad C' = C + (2K + 1).
\end{equation}
The token sequence is partitioned into $S = 32$ contiguous segments of length $n_s = \lceil L / S \rceil$, with the final segment zero-padded if necessary.

\subsection{Asymmetric Attention and Training}

\begin{figure}
\begin{center}
\includegraphics[scale=0.31]{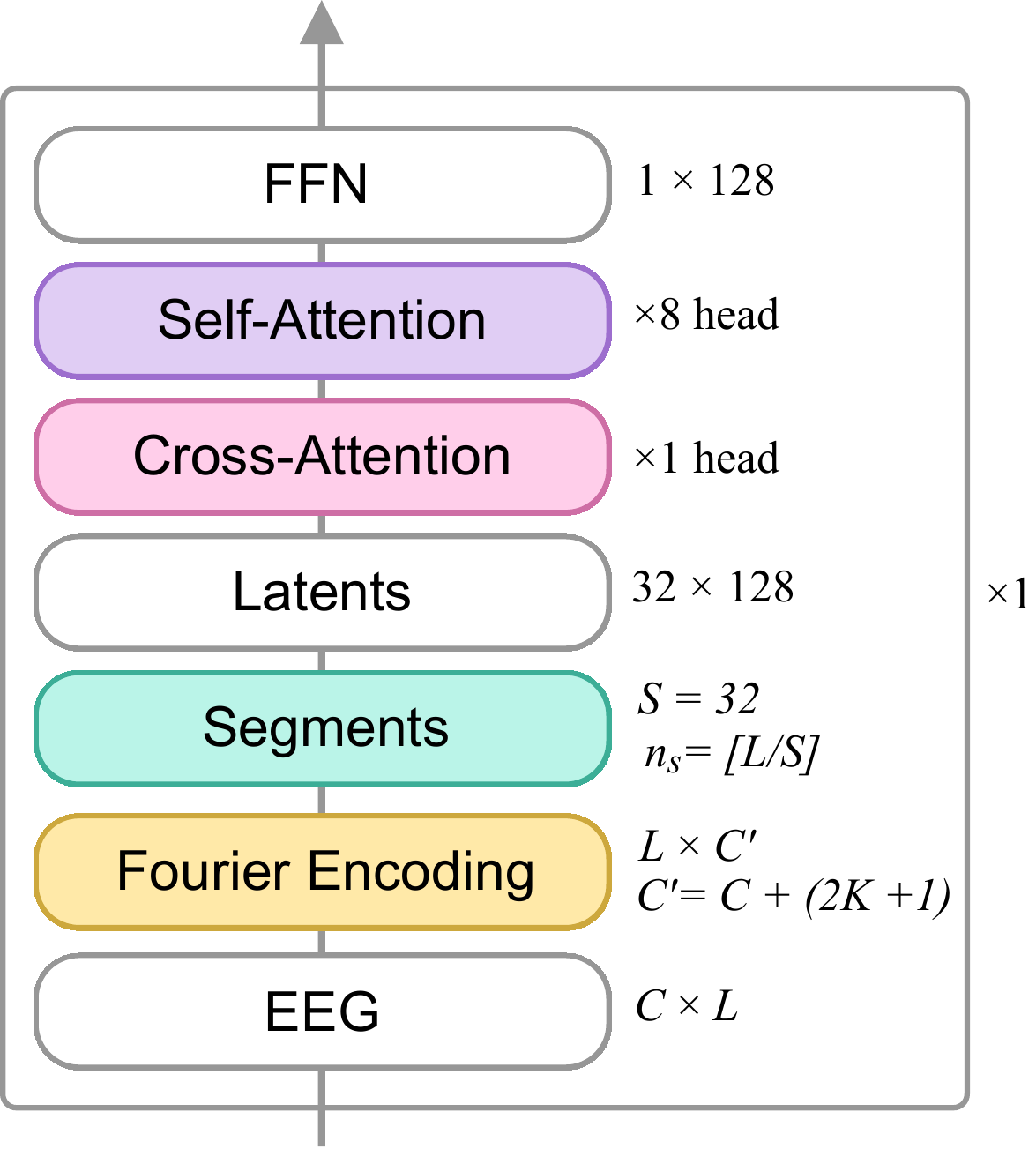}
\end{center}
\caption{Overview of the proposed EEG-based architecture. The raw EEG input with $C$ channels and $L$ temporal samples is augmented with Fourier positional encoding and divided into $S=32$ temporal segments. A cross-attention module maps each segment to a latent representation, after which self-attention and a feed-forward network model dependencies among segment-level latent states for final emotion classification.}

\label{model}
\end{figure}

A single latent state $\mathbf{e}^{(0)} \in \mathbb{R}^{d_0}$, with $d_0 = 128$, is shared across segments and updated through one layer of asymmetric attention. Each segment state aggregates information from its corresponding token subset through cross-attention:
\begin{equation}
\mathbf{e}_s = \mathbf{e}^{(0)} +
\mathrm{Attn}\!\bigl(\mathbf{e}^{(0)},\ \tilde{\mathbf{T}}_s\bigr),
\quad s = 1, \ldots, S,
\end{equation}
where $\mathbf{e}^{(0)}$ provides the query and $\tilde{\mathbf{T}}_s \in \mathbb{R}^{B \times n_s \times C'}$ provides the keys and values. The operation is asymmetric: the query is a single vector while the key-value side spans $n_s \gg 1$ tokens. All $S$ segments are processed in parallel. After cross-attention, the segment states are stacked into $\mathbf{E} \in \mathbb{R}^{B \times S \times d_0}$ and processed through $R = 8$ self-attention layers:
\begin{equation}
\mathbf{E} \leftarrow \mathbf{E} +
\mathrm{Attn}(\mathbf{E},\ \mathbf{E}).
\end{equation}
The model consists of a single processing block ($\mathrm{depth} = 1$), with the final segment states averaged across $S$ and passed through a linear classification head. 
These architectural parameters are kept fixed across all experiments, since the primary objective of this study is to investigate the contribution of EEG channels and regional electrode groups under a common modeling framework rather than to optimize the architecture itself.
Figure \ref{model} shows an overview of the architecture. 
The model is trained for $200$ epochs using AdamW with a learning rate of $10^{-4}$, a weight decay of $0.05$, and a cosine learning-rate schedule with a $20$-epoch warmup from $10^{-6}$ and a minimum learning rate of $10^{-5}$. A batch size of $32$ is employed for training, validation, and testing in the hold-out experiments, and for training and testing under LOSO evaluation.
The $3$-class task uses class-weighted cross-entropy to account for the imbalance between the classes:
\begin{equation}
w_c = \frac{1/\!\sqrt{n_c}}{\displaystyle\sum_{c'} 1/\!\sqrt{n_{c'}}},
\end{equation}
where $n_c$ is the number of training samples for class $c$. The $2$-class task uses standard cross-entropy with the mixed class excluded.
For more information and details regarding the architecture component of the specific encoder, can be found in \cite{gkikas2026edastress, gkikas2026fuse, gkikas2026unwind,gkikas_pain_icmi_2026}.

\section{Experimental Evaluation \& Results}
This section presents the experimental evaluation across two classification tasks: binary classification of positive versus negative affective states and three-class classification of positive, negative, and mixed affective states. Validation performance is reported using macro-averaged accuracy, precision, and F1 score. Test performance is reported using macro-averaged accuracy.

\subsection{Data Collection}
\label{ssec:data_collection}

In this study, the DMER dataset \cite{yang_liu_2024} was used. The dataset comprises $73$ participants aged $18$ to $35$ (mean age $23.06$, SD $3.37$); data from $7$ of the original $80$ participants were excluded due to issues with physiological signal quality. Participants watched $32$ short video clips drawn from the Stanford film library, selected through a rule-based filtering procedure and expert evaluation to elicit positive, negative, and mixed emotional states reliably: $8$ clips for positive emotions, $8$ for negative emotions, and $16$ for mixed emotions. Mixed emotion is operationally defined as the co-occurrence of positive and negative affect, validated through participant self-report on PANAS and valence-arousal-dominance scales.
EEG was recorded using the DSI-24 wireless dry-electrode system at $300$ Hz across $21$ channels, following the international $10-20$ system. This study uses EEG as the input modality. The time-aligned preprocessed EEG signals provided by DMER are used directly without additional processing. The dataset authors applied independent component analysis, bandpass filtering from 1 to $50$~Hz, $50$~Hz notch filtering, baseline removal, and re-referencing to Pz, excluding ear electrodes, yielding $18$ channels per trial. 
The DMER recordings have two durations, $20$ and $30$ seconds. In the original evaluation setting, the complete recordings are retained, and shorter trials are zero-padded to obtain a common input length. Recent analysis of DMER identified an uneven distribution of recording duration across the affective classes, indicating that the amount of padding may provide an additional class-related cue \cite{gkikas2026mupa2e}. Therefore, alongside the original padded setting, we additionally evaluate a duration-controlled setting in which all recordings are cropped to $20$ seconds, thereby removing recording length as a potential source of discriminative information while preserving the original evaluation for comparison.
Three-class classification is applied across positive, negative, and mixed affective categories. Binary classification distinguishing positive from negative affective states is also evaluated.

We employ the same stratified subject-independent partition previously reported in~\cite{gkikas2026multiscale,gkikas2026mupa2e}, comprising $48$ training, $12$ validation, and $13$ testing participants. The partition was constructed from preliminary three-class LOSO-derived subject-difficulty rankings, with the complete procedure and exact subject assignments reported in the corresponding studies. The fixed partition is retained for the extensive stride, channel-wise, and regional analyses, while LOSO is additionally employed to assess generalization across all participants.

\subsection{EEG}
\label{sec:eeg}

Table~\ref{table:eeg} reports classification performance across seven stride configurations for both tasks using all 18 EEG channels. For the $2$-class task, test accuracy ranges from $69.10\%$ at stride~$30$ to $70.09\%$ at stride~$10$, with performance remaining stable across all configurations. For the $3$-class task, test accuracy ranges from $47.51\%$ at stride~$2$ to $49.89\%$ at stride~$10$, with stride~$10$ again achieving the peak result in both tasks. The temporal stride $s$ controls the effective signal length $L = \lfloor L_0 / s \rfloor$, reducing the number of tokens and the associated computational cost as $s$ increases. Both tasks exceed their respective chance levels across all configurations ($50\%$ for $2$-class, $33.3\%$ for $3$-class).

\subsection{Channel and Regional Analysis}
\label{sec:channels}

\begin{figure*}
\begin{center}
\includegraphics[scale=0.51]{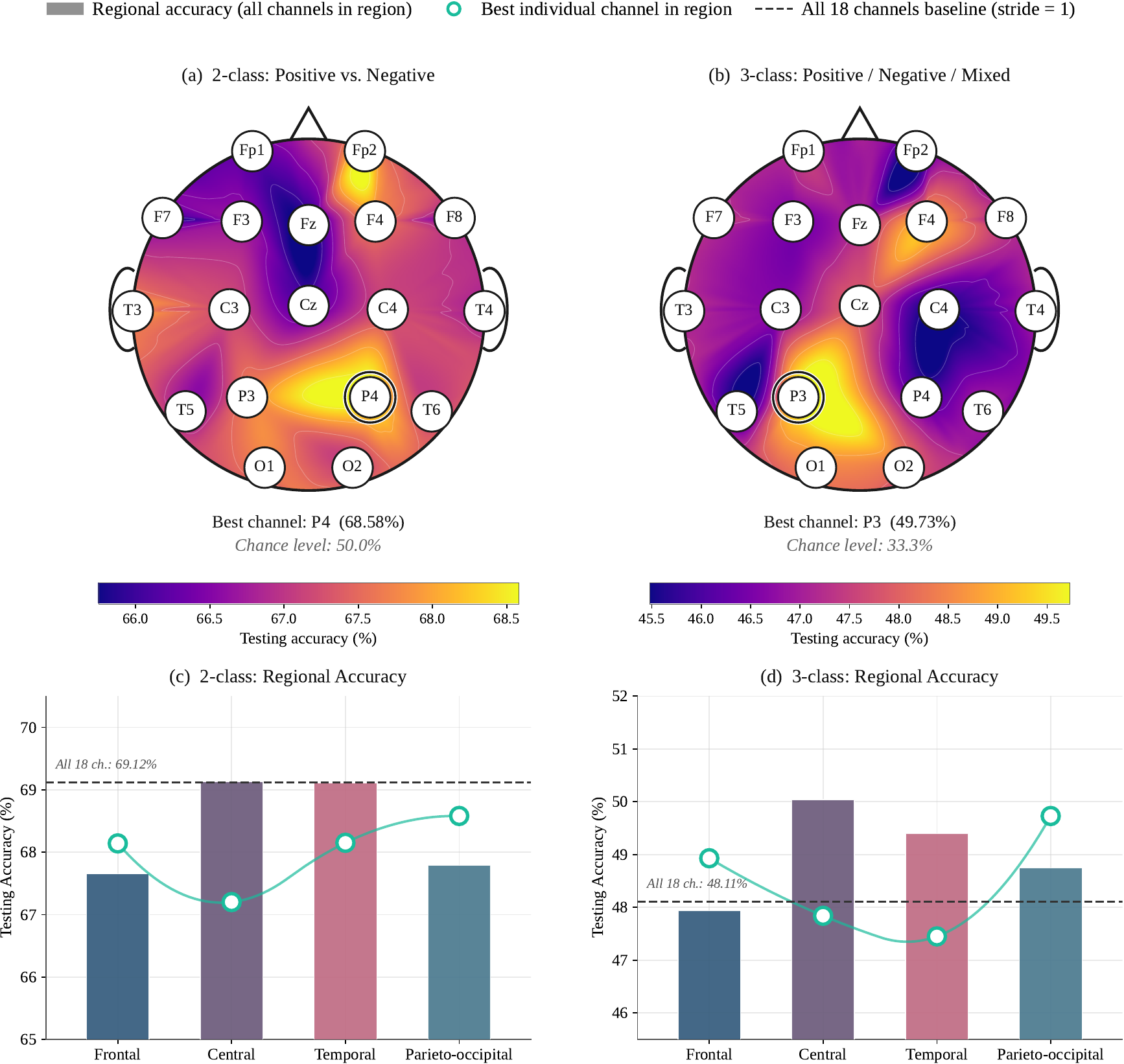}
\end{center}
\caption{EEG channel-level and regional performance analysis for the $2$-class and $3$-class emotion-recognition tasks. Panels (a) and (b) show scalp topographies of single-channel test accuracy for positive-versus-negative classification and positive/negative/mixed classification, respectively. The circled electrode indicates the best individual channel in each task. Panels (c) and (d) report regional-group test accuracy using all channels within each scalp region. Bars denote regional accuracy, circular markers denote the best individual channel within each region, and dashed horizontal lines denote the all-$18$-channel baseline at stride~$1$. The best individual channel is P4 for the $2$-class task, with $68.58\%$ accuracy, and P3 for the $3$-class task, with $49.73\%$ accuracy.}
\label{overview}
\end{figure*}

Tables~\ref{table:channels_2class} and~\ref{table:channels_3class} report per-channel and regional-group test accuracy for the $2$-class and $3$-class tasks, respectively, with all configurations evaluated at stride~$1$. Fig.~\ref{overview} visualizes the channel-level results as scalp topographies and the regional-group results as bar charts.
Individual channels perform above chance across all 18 electrodes in both tasks. For the $2$-class task, test accuracy ranges from $65.75\%$ (Fz) to $68.58\%$ (P4). For the $3$-class task, the range is $45.48\%$ (C4) to $49.73\%$ (P3), with parieto-occipital electrodes yielding the highest single-channel performance in both tasks.

Regional grouping improves over the best individual channel within each region for Central and Temporal, but not for Frontal and Parieto-occipital. For the $2$-class task, the Central ($69.13\%$) and Temporal ($69.11\%$) regions differ by only $0.01$ percentage points from the all-channels stride-1 result ($69.12\%$), while Frontal ($67.65\%$) and Parieto-occipital ($67.79\%$) show lower performance. For the $3$-class task, the Central region reaches $50.04\%$, compared with $48.11\%$ for all channels at stride~$1$ and $49.89\%$ for the highest all-channel result across strides. In the Parieto-occipital region, the best individual channel P3 ($49.73\%$) exceeds the regional group result ($48.75\%$). Given the small differences between the Central and all-channel configurations, these results are interpreted descriptively rather than as evidence of statistical superiority.
The best individual channel overall does not belong to the best-performing region in either task: P4 and P3 are parieto-occipital electrodes, while Central leads in both settings. Across both tasks, Central and Temporal consistently rank first and second among the four regions, with Frontal contributing the least. The scalp topographies in Fig.~\ref{overview}(a) and (b) show that the spatial distribution of channel-level discriminability differs between tasks: the $2$-class task exhibits a right-lateralized posterior pattern centered around C4 and P4, while the $3$-class task shows a left-posterior concentration around P3.

\begin{table}
\caption{EEG classification performance across stride configurations and tasks.}
\label{table:eeg}

\begin{center}
\begin{threeparttable}
\begin{tabular}{P{1.5cm} P{0.7cm} P{0.9cm} P{0.9cm} P{0.4cm} P{1.2cm}}
\toprule

\multicolumn{2}{c}{Input} &
\multicolumn{3}{c}{Validation} &
\multicolumn{1}{c}{Testing} \\

\cmidrule(lr){1-2}\cmidrule(lr){3-5}\cmidrule(lr){6-6}
Channel & Stride & Accuracy & Precision & F1 & Accuracy \\

\midrule
\midrule
\multicolumn{6}{l}{\textit{2-class: positive vs.\ negative}} \\
\midrule
All & 1  & 69.79 & 69.87 & 69.76 & 69.12 \\\hdashline
All & 2  & 69.27 & 69.37 & 69.23 & 69.59 \\\hdashline
All & 5  & 69.27 & 69.32 & 69.25 & 69.59 \\\hdashline
All & 10 & 69.27 & 69.37 & 69.23 & 70.09 \\\hdashline
All & 15 & 69.27 & 69.37 & 69.23 & 70.08 \\\hdashline
All & 20 & 68.75 & 68.82 & 68.72 & 69.59 \\\hdashline
All & 30 & 69.27 & 69.32 & 69.25 & 69.10 \\

\midrule
\multicolumn{6}{l}{\textit{3-class: positive, negative, mixed}} \\
\midrule
All & 1  & 46.35 & 45.76 & 45.58 & 48.11 \\\hdashline
All & 2  & 45.83 & 44.40 & 44.25 & 47.51 \\\hdashline
All & 5  & 46.53 & 46.19 & 45.07 & 48.44 \\\hdashline
All & 10 & 46.35 & 45.44 & 44.95 & 49.89 \\\hdashline
All & 15 & 46.70 & 47.59 & 40.47 & 49.23 \\\hdashline
All & 20 & 46.88 & 44.72 & 43.63 & 48.59 \\\hdashline
All & 30 & 46.88 & 45.47 & 45.29 & 47.52 \\

\bottomrule
\end{tabular}
\begin{tablenotes}[para,flushleft]
\scriptsize
\item All: all 18 EEG channels are used.
\end{tablenotes}
\end{threeparttable}
\end{center}
\end{table}

\begin{table}
\caption{EEG per-channel and regional-group classification performance for the 2-class task (positive vs.\ negative).}
\label{table:channels_2class}
\begin{center}
\begin{threeparttable}
\begin{tabular}{P{1.9cm} P{1.0cm} P{0.9cm} P{0.9cm} P{0.4cm} P{1.2cm}}
\toprule
\multicolumn{2}{c}{Input} &
\multicolumn{3}{c}{Validation} &
\multicolumn{1}{c}{Testing} \\
\cmidrule(lr){1-2}\cmidrule(lr){3-5}\cmidrule(lr){6-6}
Region & Channel & Accuracy & Precision & F1 & Accuracy \\
\midrule
\midrule
            & F3  & 68.75 & 68.88 & 68.70 & 66.69 \\\hdashline
            & F4  & 68.75 & 68.88 & 68.70 & 67.67 \\\hdashline
            & F7  & 67.71 & 67.72 & 67.70 & 66.22 \\\hdashline
            & F8  & 68.23 & 68.23 & 68.23 & 66.71 \\\hdashline
            & Fp1 & 69.27 & 69.37 & 69.32 & 66.26 \\\hdashline
            & Fp2 & 68.23 & 68.39 & 68.16 & 68.14 \\\hdashline
            & Fz  & 67.71 & 67.99 & 67.58 & 65.75 \\\hdashline
\rowcolor{mygray} Frontal     & All & 68.23 & 68.39 & 68.16 & 67.65 \\
\midrule
            & C3  & 68.23 & 68.33 & 68.19 & 67.20 \\\hdashline
            & C4  & 68.75 & 68.96 & 68.67 & 67.20 \\\hdashline
            & Cz  & 69.27 & 69.53 & 69.17 & 66.23 \\\hdashline
\rowcolor{mygray} Central     & All & 70.31 & 70.37 & 70.29 & 69.13 \\
\midrule
            & T3  & 67.71 & 67.74 & 67.69 & 68.15 \\\hdashline
            & T4  & 68.23 & 68.69 & 68.03 & 66.72 \\\hdashline
            & T5  & 67.19 & 67.74 & 66.93 & 66.72 \\\hdashline
            & T6  & 67.19 & 68.05 & 66.79 & 67.13 \\\hdashline
\rowcolor{mygray} Temporal    & All & 68.23 & 68.57 & 68.08 & 69.11 \\
\midrule
            & P3  & 67.19 & 67.28 & 67.14 & 67.65 \\\hdashline
            & P4  & 68.23 & 68.39 & 68.16 & 68.58 \\\hdashline
            & O1  & 69.27 & 69.44 & 69.20 & 67.67 \\\hdashline
            & O2  & 68.75 & 68.82 & 68.72 & 67.14 \\\hdashline
\rowcolor{mygray} Parieto-occipital & All & 68.75 & 69.05 & 68.63 & 67.79 \\
\bottomrule
\end{tabular}
\begin{tablenotes}[para,flushleft]
\scriptsize
\item \textcolor{darkgray}{Grey} rows report grouped-region results using all channels within that brain region combined.
\end{tablenotes}
\end{threeparttable}
\end{center}
\end{table}

\begin{table}
\caption{EEG per-channel and regional-group classification performance for the 3-class task (positive, negative, mixed).}
\label{table:channels_3class}
\begin{center}
\begin{threeparttable}
\begin{tabular}{P{1.9cm} P{1.0cm} P{0.9cm} P{0.9cm} P{0.4cm} P{1.2cm}}
\toprule
\multicolumn{2}{c}{Input} &
\multicolumn{3}{c}{Validation} &
\multicolumn{1}{c}{Testing} \\
\cmidrule(lr){1-2}\cmidrule(lr){3-5}\cmidrule(lr){6-6}
Region & Channel & Accuracy & Precision & F1 & Accuracy \\
\midrule
\midrule
            & F3  & 44.44 & 33.77 & 38.34 & 46.52 \\\hdashline
            & F4  & 43.58 & 32.96 & 37.51 & 48.93 \\\hdashline
            & F7  & 44.62 & 33.95 & 38.52 & 47.44 \\\hdashline
            & F8  & 45.49 & 51.98 & 41.73 & 46.98 \\\hdashline
            & Fp1 & 44.97 & 46.72 & 40.32 & 47.32 \\\hdashline
            & Fp2 & 46.88 & 45.43 & 45.40 & 45.68 \\\hdashline
            & Fz  & 44.79 & 33.59 & 38.39 & 47.18 \\\hdashline
\rowcolor{mygray} Frontal     & All & 45.83 & 44.51 & 44.56 & 47.94 \\
\midrule
            & C3  & 45.49 & 45.98 & 45.02 & 46.70 \\\hdashline
            & C4  & 46.70 & 48.91 & 46.63 & 45.48 \\\hdashline
            & Cz  & 45.66 & 35.14 & 39.61 & 47.84 \\\hdashline
\rowcolor{mygray} Central     & All & 48.44 & 46.86 & 46.31 & 50.04 \\
\midrule
            & T3  & 45.49 & 59.27 & 42.01 & 47.45 \\\hdashline
            & T4  & 44.62 & 44.10 & 44.32 & 47.18 \\\hdashline
            & T5  & 44.27 & 33.80 & 38.27 & 45.69 \\\hdashline
            & T6  & 45.66 & 46.56 & 45.95 & 47.00 \\\hdashline
\rowcolor{mygray} Temporal    & All & 47.05 & 47.39 & 44.70 & 49.40 \\
\midrule
            & P3  & 47.57 & 47.40 & 46.82 & 49.73 \\\hdashline
            & P4  & 43.92 & 32.79 & 37.55 & 46.18 \\\hdashline
            & O1  & 46.18 & 45.79 & 44.25 & 48.43 \\\hdashline
            & O2  & 46.88 & 46.88 & 45.89 & 48.08 \\\hdashline
\rowcolor{mygray} Parieto-occipital & All & 46.01 & 43.94 & 39.50 & 48.75 \\
\bottomrule
\end{tabular}
\begin{tablenotes}[para,flushleft]
\scriptsize

\end{tablenotes}
\end{threeparttable}
\end{center}
\end{table}

\subsection{Comparison with Established EEG Architectures}

Although benchmarking is not the primary objective of this study, comparisons with established EEG architectures are included to contextualize the obtained classification performance. 
The main emphasis of this study remains on characterizing how emotion-related EEG information is distributed across individual electrodes and regional electrode groups.
For a consistent comparison, EEGNet, ShallowConvNet, EEG Conformer, and DGCNN were re-run using the same subject partitions, signal-length settings, channel configurations, and training and evaluation protocol as the proposed framework. For the 2-class task, all architectures employ all $18$ EEG channels with a temporal stride of $10$, corresponding to the highest-performing configuration of the proposed framework. For the 3-class task, all architectures utilize the Central electrode group (C3, C4, Cz) with a temporal stride of $1$, which achieved the strongest performance during the regional analysis. Both the original padded recordings and the duration-controlled $20$-second recordings are evaluated under the stratified hold-out and LOSO protocols.

\begin{table*}
\caption{Comparison with established EEG architectures under the original padded and duration-controlled cropped settings for the 2-class and 3-class emotion-recognition tasks.}
\label{tab:baseline_comparison}

\begin{center}
\begin{threeparttable}
\begin{tabular}{P{0.8cm} P{1.2cm} P{1.5cm} P{2.25cm} P{2.25cm} P{2.25cm} P{2.25cm} P{2.25cm}}
\toprule

\multirow[c]{3}{*}{Task} &
\multirow[c]{3}{*}{Samples} &
\multirow[c]{3}{*}{Protocol} &
\multicolumn{5}{c}{Testing Accuracy (\%)} \\

\cmidrule(lr){4-8}
& & & Proposed & EEGNet & ShallowConvNet & EEG Conformer & DGCNN \\

\midrule
\midrule

\multirow[c]{4}{*}{2-class}
& Pad  & Hold-out & 70.09 & 68.14 & \textbf{70.12} & 68.12 & 62.24 \\\cdashline{2-8}
& Crop & Hold-out & 56.41 & 56.62 & 59.27 & \textbf{60.04} & 57.00 \\\cdashline{2-8}
& Pad  & LOSO & 74.73$\pm$4.54 & 70.69$\pm$5.22 & \textbf{77.56$\pm$6.10} & 76.36$\pm$6.64 & 70.98$\pm$9.61 \\\cdashline{2-8}
& Crop & LOSO & \textbf{73.97$\pm$9.08} & 64.98$\pm$9.41 & 70.00$\pm$9.69 & 71.65$\pm$9.71 & 69.17$\pm$11.23 \\

\midrule

\multirow[c]{4}{*}{3-class}
& Pad  & Hold-out & \textbf{50.04} & 43.27 & 47.18 & 44.72 & 34.47 \\\cdashline{2-8}
& Crop & Hold-out & 39.32 & 38.00 & \textbf{40.59} & 38.00 & 34.14 \\\cdashline{2-8}
& Pad  & LOSO & 54.61$\pm$5.60 & 53.78$\pm$6.10 & \textbf{57.64$\pm$5.48} & 56.54$\pm$6.25 & 43.47$\pm$6.23 \\\cdashline{2-8}
& Crop & LOSO & 48.40$\pm$6.86 & 45.96$\pm$6.51 & \textbf{51.82$\pm$6.88} & 51.45$\pm$6.45 & 40.33$\pm$5.75 \\

\bottomrule
\end{tabular}

\begin{tablenotes}[para,flushleft]
\scriptsize
\item Hold-out values denote test macro-averaged accuracy. LOSO values denote mean test macro-averaged accuracy $\pm$ standard deviation across held-out participants. For the 2-class task, all architectures employ all 18 EEG channels with temporal stride $10$. For the 3-class task, all architectures employ the Central electrode group (C3, C4, Cz) with temporal stride $1$. Pad denotes the original variable-duration recordings with shorter trials zero-padded to a common length; Crop denotes the duration-controlled setting in which all recordings are cropped to $20$ seconds. Bold indicates the highest accuracy within each experimental setting.
\end{tablenotes}

\end{threeparttable}
\end{center}
\end{table*}

Table~\ref{tab:baseline_comparison} compares the proposed framework with the established EEG architectures under the padded and duration-controlled settings. Under hold-out evaluation, the proposed framework achieves the highest padded 3-class accuracy of $50.04\%$, while ShallowConvNet and EEG Conformer provide the strongest results in several remaining settings. Under LOSO evaluation, ShallowConvNet achieves the highest padded accuracies of $77.56\%$ and $57.64\%$ for the 2-class and 3-class tasks, respectively, whereas the proposed framework reaches the highest cropped 2-class accuracy of $73.97\%$. For the proposed framework, the 3-class LOSO evaluation additionally yields macro F1 scores of $51.61\pm6.99\%$ and $45.22\pm8.67\%$ for the padded and cropped settings, respectively, providing a class-balanced assessment across positive, negative, and mixed affective states.

For additional context, recent studies have evaluated the same DMER dataset using the same subject-independent partition employed in this work. The authors in~\cite{gkikas2026multiscale} employed the same stratified hold-out partition with duration-controlled $20$-second EEG recordings, achieving $65.22\%$ and $45.43\%$ test accuracy for the 2-class and 3-class tasks, respectively. Additionally, MUPA\textsuperscript{2}E~\cite{gkikas2026mupa2e} investigated the effect of recording duration on DMER, reporting $70.07\%$ accuracy using the original padded recordings and $62.71\%$ after cropping all recordings to $20$ seconds in the binary multimodal setting, while the cropped configuration reached $76.37\%$ under LOSO evaluation.


\subsection{Overall Analysis \& Discussion}

All-channel EEG classification exceeds chance in both tasks across all stride configurations. The 2-class task reaches $70.09\%$ test accuracy at stride~$10$, $20.09$ percentage points above the $50\%$ chance level. The 3-class task reaches $49.89\%$, $16.56$ percentage points above the $33.3\%$ chance level, with the lower absolute margin reflecting the additional difficulty introduced by the mixed emotion class. The $3$-class result of $49.89\%$ exceeds the EEG-only leave-one-subject-out baseline of $44.16\%$ established during the stratified split protocol design, demonstrating that the deep learning approach improves over classical machine learning under a subject-independent evaluation.
Performance across stride configurations is more stable for the $2$-class task (range $69.10-70.09\%$) than for the $3$-class task (range $47.51-49.89\%$), suggesting the framework is more robust to temporal sampling when classifying pure affective states and more sensitive to input temporal context when mixed emotions are included.
The regional analysis shows that three central electrodes---C3, C4, and Cz---provide performance comparable to the full $18$-channel configuration. In the $2$-class task, the Central result differs from the all-channels stride-1 result by only $0.01$ percentage points ($69.13\%$ vs.\ $69.12\%$), while in the $3$-class task it reaches $50.04\%$ compared with the highest all-channels result of $49.89\%$. Given the small magnitude of these differences, they are interpreted descriptively rather than as evidence of statistical superiority. Nevertheless, the comparable performance obtained using only $3$ electrodes suggests that reduced-channel configurations warrant further investigation for lightweight EEG systems.

The relationship between individual channel performance and regional group performance varies across regions. In the Central and Temporal regions, combining channels produces consistent gains: the Central group outperforms its best individual channel by $1.93$ percentage points in the $2$-class task and by $2.20$ percentage points in the $3$-class task. In the Frontal and Parieto-occipital regions, the group does not consistently improve over the best individual channel. For the 3-class task, P3 alone ($49.73\%$) exceeds the Parieto-occipital group result ($48.75\%$), indicating that other channels within that region introduce variability that reduces the group benefit. These patterns suggest that combining regional channels is most effective in central and temporal areas. Although the channel-wise and regional analyses provide spatial information regarding which electrode subsets retain discriminative performance, they do not explain the internal decision process of the deep learning model. Future work should therefore incorporate model-level explainability methods, such as gradient-based attribution or attention analysis, to investigate how temporal EEG information contributes to individual predictions.
A further limitation is that the present evaluation is restricted to the DMER dataset, which was collected under controlled experimental conditions. Consequently, the generalization of the observed channel and regional patterns to other populations, acquisition systems, and naturalistic affective settings remains unknown. Future work should therefore evaluate the framework across additional datasets and acquisition conditions to examine the robustness of these spatial EEG findings.

The scalp topographies reveal different spatial distributions of discriminative information between the two tasks. 
The $2$-class task shows a right-lateralized posterior pattern, with the highest individual-channel performance concentrated around C4 and P4. The $3$-class task shows a left-posterior concentration around P3, with additional activity in the right-frontal area. The shift in lateralization across tasks suggests that binary valence discrimination and three-class mixed affective state recognition engage partially distinct spatial patterns.

The $3$-class setting, which requires discrimination among positive, negative, and mixed affective states, remains more challenging than the binary positive-versus-negative setting. In addition to the overall accuracy, the LOSO evaluation yields macro F1 scores of $51.61\pm6.99\%$ under the padded setting and $45.22\pm8.67\%$ after duration-controlled cropping, with each class contributing equally to the metric. These results provide a class-balanced characterization of subject-independent $3$-class recognition, while class-specific conclusions regarding the mixed category would require dedicated per-class evaluation.
Recognizing mixed affective states from a reduced number of EEG channels may support the development of future mental health monitoring systems with lower sensing and input-dimensionality requirements. However, the present study does not evaluate hardware deployment, inference latency, or energy consumption; therefore, the computational benefits and feasibility of real-time wearable operation require dedicated implementation-level evaluation.

\section{Conclusion}

This paper presented an EEG-based framework for mixed-emotion recognition. It evaluated the contributions of individual channels and regional groupings in binary and three-class classification settings on the DMER dataset. Both tasks exceeded their respective chance levels across all stride configurations, with the best results reaching $70.09\%$ for the $2$-class task and $49.89\%$ for the $3$-class task. These results establish a deep learning reference point for EEG-based mixed emotion classification under a subject-independent evaluation protocol.
The central electrode group (C3, C4, Cz) produced the strongest regional performance in both tasks while remaining comparable to the full $18$-channel configuration.
Because these electrodes overlie sensorimotor-related scalp areas, this pattern may reflect embodied and somatic components of emotional processing. For the $3$-class task, the best individual channel was P3, a left parietal electrode within the parieto-occipital group, whereas the $2$-class task favored P4, a right parietal electrode within the same group. This lateralization shift suggests that including mixed emotions changes the spatial pattern of EEG discriminability, potentially reflecting the additional attentional, evaluative, or integrative processing required when positive and negative affective components co-occur.
These findings motivate further investigation of reduced-channel EEG configurations for future mental health monitoring applications. In particular, the comparable performance obtained using the Central electrode group suggests that affective-state recognition can be investigated with substantially fewer electrodes than the full $18$-channel configuration. 

We suggest that future work needs to focus on explicit channel optimization for minimal wearable deployments, and also on evaluating the methods with regard to real-time EEG monitoring, something that could be beneficial for passive health assessment in several real-world scenarios \cite{montiel_cruz_gkikas_eeite_2026, gkikas_tiny_2025}.

\section*{Acknowledgments}
The authors used large language model (LLM)-based tools for language editing and improvement. All scientific content, results, and conclusions are solely the work of the authors.


\bibliographystyle{IEEEtran}
\bibliography{library}

\end{document}